\documentclass{article}
\usepackage{graphics}
\usepackage{epsfig}
\usepackage{amssymb}

\newenvironment{myeq}[1]{\begin{equation} \hspace{#1}}{\end{equation}}
\newenvironment{myeqn}[1]{\begin{displaymath} \hspace{#1}}{\end{myeq}}
\newcommand{\next}[1]{\end{displaymath} \vspace{-0.5cm} \\ \begin{displaymath} \hspace{#1}}
\newcommand{\last}[1]{\end{displaymath} \vspace{-0.5cm} \\ \begin{myeq}{#1}}
\newcommand{\more}[1]{\end{myeq}  \vspace{-0.5cm} \\ \begin{myeq}{#1}}
\newcommand{\Op}{\mathcal{O}}

\begin{document}

\begin{center}
        THE $\eta$ DECAY CONSTANT IN `RESUMMED' CHIRAL PERTURBATION THEORY \\
\end{center}
\begin{center}
        MARI\'{A}N KOLES\'{A}R and JI\v{R}\'{I} NOVOTN\'{Y} \\
\end{center}
\begin{center}
        \textsl{Institute of particle and nuclear physics, MFF UK, Prague, Czech Republic} \\
\end{center}
\begin{center}
        \textsl{E-mail address: kolesar@ipnp.troja.mff.cuni.cz, novotny@ipnp.troja.mff.cuni.cz} 
        \\
\end{center}

The recently developed 'Resummed' $\chi$PT is illustrated on the case of pseudo\-scalar meson decay constants. We try to get an estimate of the $\eta$ decay constant, which is not well known from experiments, while using several ways including the Generalized $\chi$PT Lagrangian to gather information beyond Standard next-to-leading order. We compare the results to published $\chi$PT predictions, our own Standard $\chi$PT calculations and available phenomenological estimates.
\\

\\ \vspace{-0.25cm}

\normalsize

\section{Introduction}

As was discussed recently
\cite{SternRes,Stern2,Stern1,DescotesRes,Descotes}, chiral
perturbation theory \cite{G&L2,G&L3}, the low energy effective
theory of QCD with $N_f$ light quark flavors, could posses
different behavior for $N_f$=2 and $N_f$=3. As a consequence of
vacuum fluctuations of the growing number of light quark flavors,
the most important order parameters of spontaneous chiral symmetry
breaking ($SB\chi S$), namely the pseudoscalar decay constant and
the quark condensate in the chiral limit, obey paramagnetic
inequalities $F_0(N_f+1)<F_0(N_f)$ and $\Sigma(N_f+1)<\Sigma
(N_f)$ \cite{Stern1}. In particular, the fluctuations of the sea
$\overline{s}s$-pairs need not be suppressed due to the relatively
small value of the $s$-quark mass $m_s$$\lesssim$$\Lambda _{QCD}$
and could bring about a possibly significant suppression of
$F_0(3)$ and $\Sigma(3)$ w.r.t. $N_f$=2. This should manifest
itself through the OZI rule violation in the scalar sector as can
be seen from

\begin{myeq}{0cm}
        F_0(2)^2\ =\ F_0(3)^2 + 16 m_s B_0 L_4^r - 2\bar{\mu}_K + \Op(m_s^2)
\more{0cm}
				\Sigma(2)\ =\ \Sigma(3)( 1 + \frac{32 m_s B_0}{F_0^2} L_6^r - 2\bar{\mu}_K -
                                		 \frac{1}{3}\bar{\mu}_{\eta}) + \Op(m_s^2),
\end{myeq}

\noindent where $B_0^2=\Sigma(3)/F_0(3)^2$,
$\bar{\mu}_P=\mu_P|_{m_{u,d}\to0}$. Indeed, $L_4$ and $L_6$ are
the $1/N_c$ suppressed LEC's (connected to the scalar mesons),
traditionally considered negligible. Predictions for $L_4$ and
$L_6$ derived from sum rules involving scalars
\cite{Moussallam1,Moussallam2,Descotes,Stern2}, calculations on
the lattice \cite{Aubin,MILC2006} and NNLO S$\chi$PT
\cite{BijnL4L6} produce numbers significantly different from
traditional expectations \cite{G&L3}. Convenient parameters
relating the order parameters to physical quantities can be
introduced: $Z(N_f)=F_0(N_f)^2/F_{\pi}^2$ and
$X(N_f)=2\hat{m}\Sigma(N_f)/F_{\pi}^2M_{\pi}^2$, with
$\hat{m}$=$(m_u+m_d)/2$. The large $\overline{s}s$ vacuum
fluctuations could lead to $Z(3)\ll Z(2)$, $X(3)\ll X(2)$.
Analysis \cite{SternExp} of the $K_{e_4}$ decay experimental
results for the $\pi\pi$ $s$-wave scattering length \cite{Pislak}
lead to values $X(2)$=$0.81\pm0.07$, $Z(2)$=$0.89\pm0.03$.
$\pi\pi$ and $\pi K$ scattering data constrain the three flavor
parameters much less strictly \cite{SternRes,DescotesRes},
$X(3)$$<$0.8, $Z(3)$$\sim$0.2-0.9, $Y$=$X(3)/Z(3)$$<$1.1,
$r$=$m_s/\hat{m}$$>$15. Sum rule approaches \cite{Descotes,Stern2}
yield approximately $X(2)$,$Z(2)$$\sim$0.9 and
$X(3)$,$Z(3)$$\sim$0.5 at $r$=25.

Small $X(3)$ and $Z(3)$ would lead to irregularities of the chiral
expansion connected to numerical competition of the LO and NLO
terms, which could be consequently seen as unusually large higher
order corrections. An alternative approach, dubbed `Resummed'
$\chi$PT (R$\chi$PT), has been introduced recently
\cite{SternRes}. It takes this possible scenario into account and
is based on the effective resummation of the vacuum fluctuation
discussed above. The goal of the article is to illustrate the
`Resummed' approach on the sector of decay constants and to try to
use it for theoretical predictions of the $\eta$ decay constant
and related parameters, including uncertainty estimates.

The $SU(3)_L\times SU(3)_R$ $\eta$ decay constant in the isospin limit

\begin{myeq}{0cm}
        i p_{\mu}\, F_{\eta}\ =\ \langle\,0\,|\,A_{\mu}^8\,|\,\eta(p)\,\rangle,
\end{myeq}

\noindent where $A_{\mu}^i$ are the QCD axial vector currents, can
be  calculated in $SU(3)_L\times SU(3)_R$ $\chi$PT without the
introduction of the $\eta'$ meson. In the usually investigated
$\eta$-$\eta'$ mixing sector, the following definitions are used

\begin{myeq}{0cm}
        i p_{\mu}\, F^{8,0}_{\eta,\eta'}\ =\ 
        \langle\,0\,|\,A_{\mu}^{8,0}\,|\,\eta,\eta'\,\rangle.
\end{myeq}

\noindent As can be seen, the $SU(3)_L$$\times$$SU(3)_R$ constant $F_{\eta}$ is defined identically to $F_{\eta}^8$ in the $U(3)_L$$\times$$U(3)_R$ framework.
A general two angle mixing scheme \cite{LeutwylerN,Feldmann1}

\begin{myeq}{0.5cm}
        F_{\eta}^8\, =\, F_8\cos\vartheta_8,\ F_{\eta'}^8\, =\, F_8\sin\vartheta_8,\
        F_{\eta}^0\, =\, -F_0\sin\vartheta_0,\ F_{\eta'}^0\, =\, F_0\cos\vartheta_0
\end{myeq}

\noindent has been shown to provide better agreement with experimental data and $\chi$PT predictions \cite{Feldmann1,EscFrere1,EscFrere2} than a single mixing angle scenario. \mbox{Table \ref{FEtaPhen}} collects some recent two angle phenomenological analyses leading to a value of $F_{\eta}^8$. Older one angle mixing scheme results generally provided much lower numbers $F_{\eta}^8\sim F_{\pi}$.

\begin{table} \small
\hspace{0.75cm}
\begin{tabular}{|c|c|c|@{\,}|c|c|@{\,}|c|}
        \hline \rule[-0.2cm]{0cm}{0.7cm}
                year & cit. & model input & $F_8$ & $\vartheta_8$ & $F_{\eta}^8$ \\
        \hline \rule[-0.2cm]{0cm}{0.7cm}
                2005 & \cite{EscFrere2} & & $(1.51\pm0.05)F_{\pi}$ & $(-24\pm1.6)^o$ & 
                $1.38F_{\pi}$ \\
        \hline \rule[-0.2cm]{0cm}{0.7cm}
                2000 & \cite{Pennington} & sum rules & $1.44F_{\pi}$ & $-8.4^o$ & $1.42F_{\pi}$ 
                \\
        \hline \rule[-0.2cm]{0cm}{0.7cm}
                1999 & \cite{Benayoun} & VMD & & & $1.27F_{\pi}$ \\
        \hline \rule[-0.2cm]{0cm}{0.7cm}
                1998 & \cite{Feldmann2} & & $1.26F_{\pi}$ & $-21.2^o$ & $1.17F_{\pi}$ \\
        \hline
\end{tabular}
\label{FEtaPhen}
\caption{Recent two angle $\eta$-$\eta'$ analyses leading to a value of $F_{\eta}^8$}
\end{table} \normalsize

Several recent $\chi$PT results can be cited. Standard $\chi$PT to $\Op(p^6)$ \cite{Bijnens2} gives $F_{\eta}/F_{\pi} = 1 + 0.242 + 0.066 = 1.308.$ Large $N_c$ $\chi$PT to NNLO \cite{K&L1} leads to $F_8=1.34F_{\pi}$, $\vartheta=-22^o$ and thus $F_{\eta}^8 = 1.24F_{\pi}$. We build on the `Resummed' $\chi$PT result \cite{SternRes} $F_{\eta}^2 = F_{\pi}^2(1.651+0.036Y)$ (at $r$=24, remainders neglected).

\section{\boldmath Decay constants in `Resummed' $\chi$PT}

`Resummed' $\chi$PT \cite{SternRes} starts from the same form of the effective Lagrangian as the Standard variant (S$\chi$PT) \cite{G&L3}. The difference is in the treatment of the chiral series, R$\chi$PT assumes possible irregularities. Overall convergence to all orders is taken for granted, but only for expansions directly obtained from the generating functional. These `bare' expansions are then dealt with additional caution.

The first step is to derive a strict chiral expansion fully expressed in terms of the original parameters of the effective Lagrangian. In our case we have

\begin{myeq}{-3cm} \label{FPi}
        F_{\pi}^2\, =\, F_0^2( 1-4\mu_{\pi}-2\mu_K ) + 16B_0\hat{m}(L_4(r+2)+L_5) +
                        \Delta_{F_{\pi}}^{(4)}
\more{0cm} \label{FK}
        F_K^2\, =\, F_0^2(1-\frac{3}{2}\mu_{\pi}-3\mu_K-\frac{3}{2}\mu_{\eta}) +
                    16B_0\hat{m}(L_4(r+2)+\frac{1}{2}L_5(r+1)) + \Delta_{F_K}^{(4)}
\more{-2.25cm} \label{FEta}
        F_{\eta}^2\ =\ F_0^2(1-6\mu_K) + 16B_0\hat{m}(L_4(r+2)+\frac{1}{3}L_5(2r+1)) +
                       \Delta_{F_{\eta}}^{(4)}\ .
\end{myeq}

\noindent The expansions for the squares of the decay constants
are used, as  they are directly related to two point Green
functions obtained from the generating functional. At this point,
the chiral logs $\mu_P=m_P^2/32\pi^2F_0^2 \ln(m_P^2/\mu^2)$
contain non-physical $\Op(p^2)$ masses $m_{\pi}^2=2B_0\hat{m}$,
$m_K^2=B_0\hat{m}(1+r)$, $m_{\eta}^2=2/3B_0\hat{m}(1+2r)$.
$\Delta_{F_P}^{(4)}$ denote the higher order remainders, not
neglected in this approach.

The second step is the definition of the bare expansion, which usually involves changes to the strict form in order to incorporate additional requirements, such as physically correct analytical structure. In our case this narrows down to a question, whether to replace the original leading order masses inside the chiral logarithms with physical ones. In some cases (see \cite{EtaPi}) this is a nontrivial question, so we will keep both options and evaluate them.

The next stage is the reparametrization of the unknown LEC's in terms of physical observables. In R$\chi$PT the leading order ones are left free, only re-expressed in terms of more convenient parameters $r,Z$ and $X$ resp. $Y$. Two NLO LEC's are present in our formulae, the equations for $F_{\pi}$ and $F_K$ (\ref{FPi},\ref{FK}) can be used for the reparametrization. Note that this is done in a pure algebraic way, no additional expansion is made. The final formula for the $\eta$ decay constant \cite{SternRes} is then obtained by insertion into (\ref{FEta})

\begin{myeq}{0cm}
        F_{\eta}^2 = \frac{1}{3}\Big[4F_K^2-F_{\pi}^2 + \frac{M_{\pi}^2 Y}{16\pi^2}
                      (\ln \frac{m_{\pi}^2}{m_K^2} + (2r+1)\ln \frac{m_{\eta}^2}{m_K^2})
                      + 3\Delta_{F_{\eta}}^{(4)} - 4\Delta_{F_K}^{(4)} + 
                      \Delta_{F_{\pi}}^{(4)}\Big].
\end{myeq}

\noindent The expression is valid to all orders, it's only divided into an explicitly calculated part and the unknown higher order remainders $\Delta_{F_P}^{(4)}$.

The last step consists of the treatment of the remainders. We will use three ways to estimate them. The first relies on an assumption about the convergence of the chiral series \cite{SternRes, DescotesRes} and assumes the typical size of the NNLO remainders is $|\Delta_{F_P}^{(4)}|\sim 0.1 F_P^2$. These are added in squares to obtain the final uncertainty. The result is a prediction in the sense that a value significantly outside of the resulting variance is not compatible with such an assumption about a reasonably quick convergence of the chiral expansion.

Then we try to use information outside core $\chi$PT to get a feeling about remainder magnitudes. As can be seen, the R$\chi$PT framework is very suitable for incorporating such additional sources of information. We collect various published estimates for $L_5^r$ and use them to check the remainder differences $\Delta_{F_K}^{(4)}$-$\Delta_{F_{\pi}}^{(4)}$ and $\Delta_{F_{\eta}}^{(4)}$-$\Delta_{F_{\pi}}^{(4)}$. We also use the Generalized $\chi$PT Lagrangian \cite{GCHPT,SternPiPi} to get a sense of the magnitude of the higher order corrections

\begin{myeq}{0cm}
        \Delta_{F_P}^{(4)}\ =\ {(F_P^{(2)})}_{G\chi PT}-F_P^{(2)}+\Delta_{F_P}^{(G\chi PT)}.
\end{myeq}

\noindent More details about this procedure will be published elsewhere \cite{EtaPi}.

\section{\boldmath Numerical results}

For the numerical results we use the physical values $M_{\pi}$=135MeV, $M_K$=496MeV, $M_{\eta}$=548MeV, $M_{\rho}$=770MeV, $F_{\pi}$=92.4MeV and $F_K$=113MeV.
At first, let us investigate the NLO Standard $\chi$PT. There are several differences compared to the  procedure outlined in the previous section. One can use the quadratic form of the expansion obtained from the two point Green function or a linearized form, as is more usual. For the LEC reparametrization inverted expansions for $F_0^2$ and $2B_0\hat{m}$ are used, while $r$ is fixed at $r=r_2=2M_K^2/M_{\pi}^2-1$ or $r=\tilde{r}_2=3M_{\eta}^2/2M_{\pi}^2-1/2$. One then obtains the following formulae

\begin{myeq}{-0.25cm}
        \frac{F_{\eta}}{F_{\pi}} = 1+2\mu_{\pi}-2\mu_K +
                     \frac{8M_{\pi}^2(r-1)}{3F_{\pi}^2}L_5^r,\ \
        \frac{F_{\eta}^2}{F_{\pi}^2} = 1+4\mu_{\pi}-4\mu_K +
                     \frac{16M_{\pi}^2(r-1)}{3F_{\pi}^2}L_5^r
\end{myeq}

\noindent where the chiral logs contain physical masses only $\mu_P=M_P^2/32\pi^2F_{\pi}^2 \ln(M_P^2/\mu^2)$.

As for $L_5^r$, we opted to use the published values
$L_5^r(M_{\rho})=(1.4\pm0.5).10^{-3}$ ($\Op(p^4)$ fit \cite{G&L3,LEC4}) and $L_5^r(M_{\rho})=(0.65\pm0.12).10^{-3}$ ($\Op(p^6)$ fit \cite{LEC6}).

All these possibilities differ merely in redefinitions of the usually neglected remainders. Table \ref{SCHPTTable} shows that it might be worth to spend the additional effort to bring the higher order uncertainties explicitly under control. Numerically, the sensitivity to the change in $L_5^r$ is in the range $\Delta F_{\eta}/F_{\pi}=(0.11-0.14)\Delta L_5.10^3$.

\begin{table} \small
\begin{tabular}{|c|@{\,}|c|c|c|c|}
        \hline \rule[-0.2cm]{0cm}{0.7cm}
                     expansion & $O(p^4)\,L_5$, $r$=$r_2$ & $O(p^6)\,L_5$, $r$=$r_2$ &
                     $O(p^4)\,L_5$, $r$=$\tilde{r}_2$ & $O(p^6)\,L_5$, $r$=$\tilde{r}_2$ \\
        \hline \rule[-0.2cm]{0cm}{0.7cm}
                     $F_{\eta}$ & \textbf{1.31$\pm$0.07} & 1.21$\pm$0.02 & 1.29$\pm$0.07 &
                     1.19$\pm$0.02 \\
        \hline \rule[-0.2cm]{0cm}{0.7cm}
                     $\surd F_{\eta}^2$ & 1.27$\pm$0.06 & 1.19$\pm$0.01 & 1.25$\pm$0.05 &
                 \textbf{1.17$\pm$0.01} \\
        \hline
\end{tabular}
\caption{Various NLO S$\chi$PT results for the $\eta$ decay constant in $F_{\pi}$ units.}
\label{SCHPTTable}
\end{table} \normalsize

Proceeding to R$\chi$PT, we generally investigate a standard and a low $r$
scenario $r\sim$15-25 and vary $Y$ in the range 0-1.6. Keep in mind, though, that the
$\pi\pi$ and $\pi K$ scattering analyses \cite{SternRes,DescotesRes} suggest $Y<1.1$.

Let us first neglect the remainders and have a look on the dependence of
the explicitly calculated part on the free parameters $Y,r$ and the treatment of chiral logs. As can be seen from Fig.1, the dependence on both is very small. For physical masses inside the logs one gets $F_{\eta}^2/F_{\pi}^2=1.661-0.011Y+0.002Y r$. The decay constant sector might thus be insensitive to the particular scenario of $SB\chi S$ and more information is needed to extract the values of the parameters.

Neglecting these weak dependencies one gets the sensitivity on the remainders as $\Delta F_{\eta}/F_{\pi}=1.5\cdot 10^{-5} \surd( (3\Delta_{F_{\eta}}^{(4)})^2 + (4\Delta_{F_K}^{(4)})^2 + (\Delta_{F_{\pi}}^{(4)})^2 )$.

\vspace{0.4cm}
\noindent
\begin{minipage}{0.6\textwidth}
        \epsfig{figure=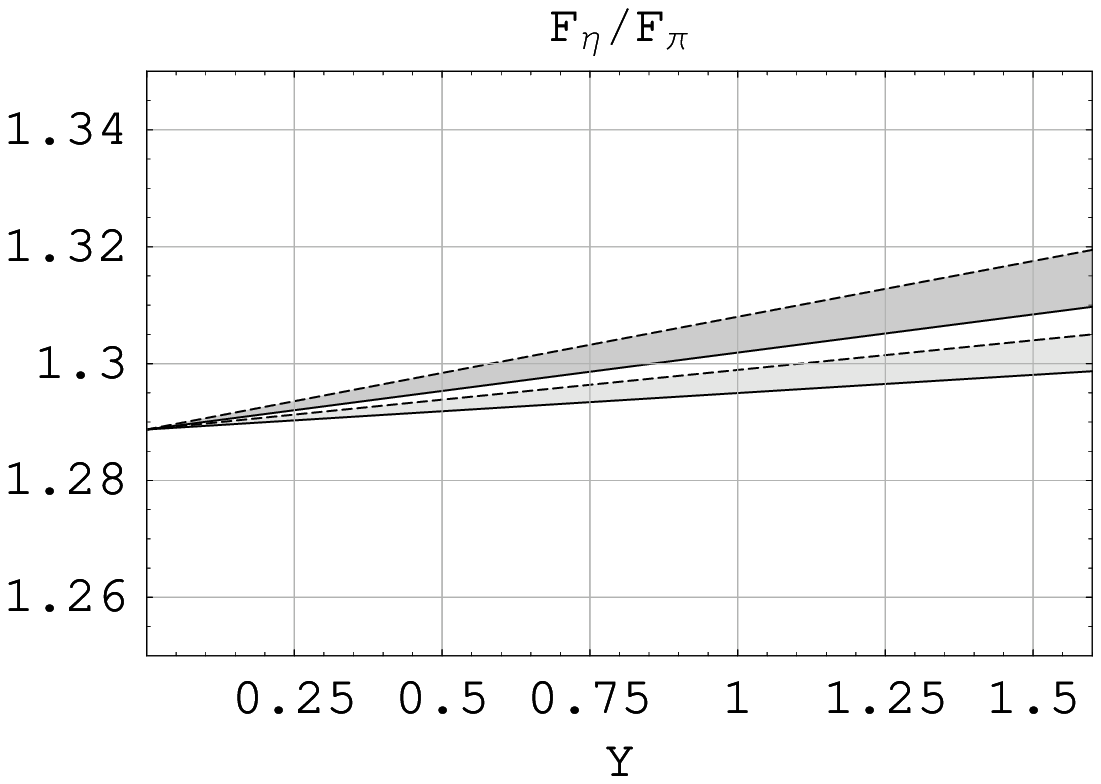,width=\textwidth} \\
        Figure 1: $F_{\eta}$ in R$\chi$PT, remainders neglected.
        					Chiral logs: solid - physical masses, dashed - $\Op(p^2)$ masses.
        					Dark: $r=25$, light: $r=15$
\end{minipage}
\begin{flushright} \begin{minipage}{0.38\textwidth} \vspace{-6.5cm}
        Applying the 10\% uncertainty remainder estimate, the following all order approximation 
        is obtained
        \begin{myeq}{0.5cm}
            F_{\eta}\ =\ (1.3\pm0.1)F_{\pi}.
        \end{myeq}
        All phenomenological and theoretical results cited in the introduction fall in or
        very close to this range and are thus compatible with a reasonable convergence of chiral
        series. However, the mentioned one mixing angle scheme results are significantly 
        outside.       
\end{minipage} \end{flushright}  \vspace{-0.3cm}

The difference $F_K^2$-$F_{\pi}^2$ (\ref{FPi},\ref{FK}) depends only on $L_5^r$. This yields an order estimate on the remainder difference $\Delta_{F_K}^{(4)}$-$\Delta_{F_{\pi}}^{(4)}$ if independent information on $L_5^r$ can be gathered. Several estimates for $L_5^r$ beyond $\Op(p^4)$ $\chi$PT are available:

\vspace{-0.2cm}
\begin{itemize}
        \item[-] S$\chi$PT $\Op(p^6)$ fit: $L_5^r(M_\rho)\sim(0.5-1.0)\cdot10^{-3}$
                 \cite{LEC6,Bijnens2}
                 \vspace{-0.25cm}
        \item[-] Resonance saturation: $L_5^r\sim(1.6-2.1)\cdot10^{-3}$ \cite{Moussallam2}
                 \vspace{-0.25cm}
        \item[-] QCD sum rules: $L_5^r(M_\rho)>1.0\cdot10^{-3}$ \cite{Stern2,Descotes}
                 \vspace{-0.25cm}
        \item[-] $\chi$PT on lattice: $L_5^r\sim1.8-2.2\cdot10^{-3}$ \cite{Aubin,MILC2006}
\end{itemize}
\vspace{-0.2cm}

The result of varying $L_5^r(M_{\rho})$ in the range $(0.5-2).10^{-3}$ can be seen in Fig.2.  $\Op(p^2)$ masses were kept inside logarithms, physical ones make the remainder estimate somewhat larger. The estimate is compatible with small remainders.

\begin{figure}[t]
        \hspace{-0.5cm}
        \epsfig{figure=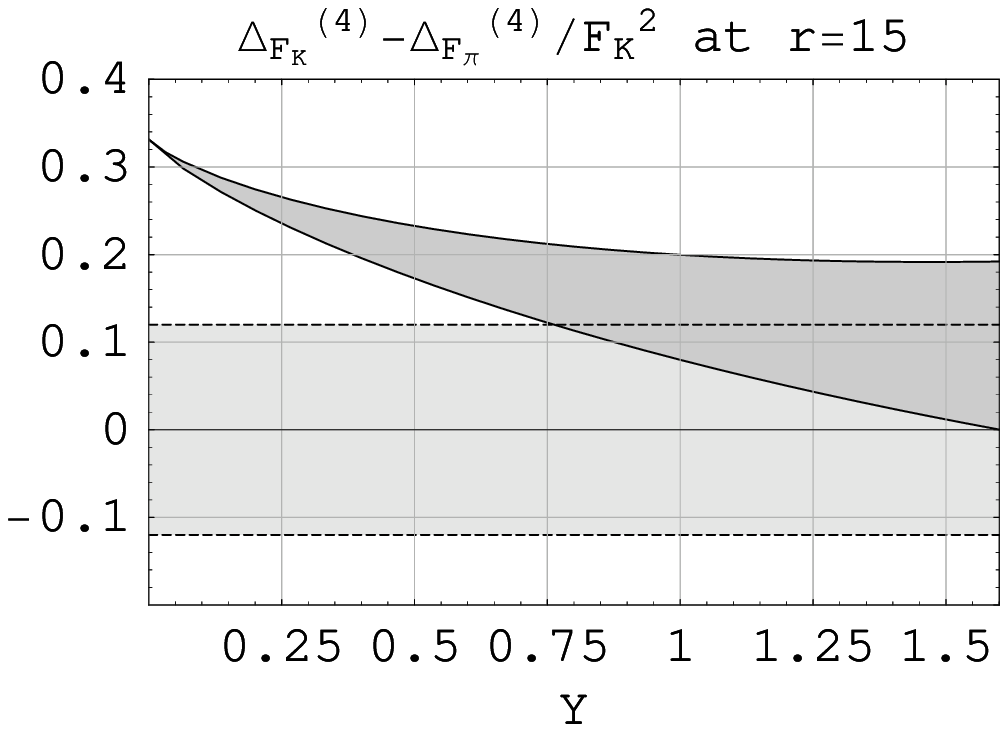,width=0.53\textwidth}
        \hspace{-0.5cm}
        \epsfig{figure=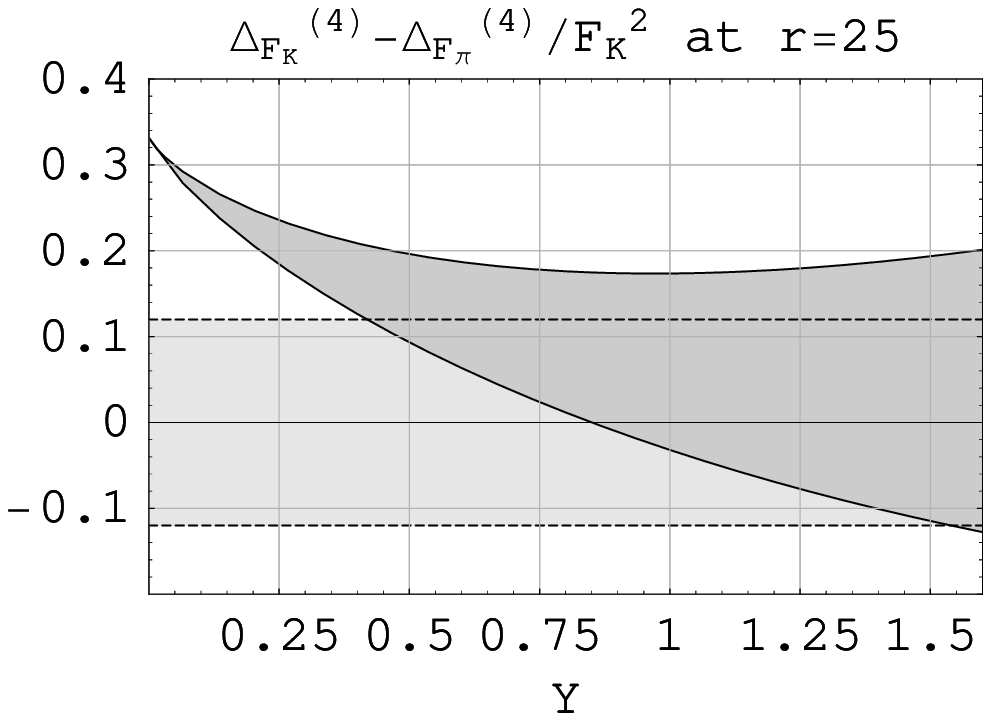,width=0.53\textwidth} \\ \\
        Figure 2: $\Delta_{F_K}^{(4)}$-$\Delta_{F_{\pi}}^{(4)}$ estimate for
        					$L_5^r(M_{\rho})\sim(0.5-2).10^{-3}$ (dark band). Upper bound corresponds to 
        					low values of $L_5^r$. Left: $r$=15, right: $r$=25. \\
      						Light: expected uncertainty $\pm0.12F_K^2$ from the 10\% uncertainty estimate.
\end{figure}

We can also utilize the information about the difference $F_{\eta}^2$-$F_{\pi}^2$ (\ref{FPi},\ref{FEta}). If we use the latest phenomenological result $F_{\eta}^8\sim1.38F_{\pi}$ \cite{EscFrere2} as an input, an estimate of $\Delta_{F_{\eta}}^{(4)}$-$\Delta_{F_{\pi}}^{(4)}$ is obtained. It should be stressed that older results produced lower values, so this should be taken as a preliminary look on the possible consequences if such a higher value of $F_{\eta}$ was confirmed. We don't make a full statistical analysis, only provide some first feelings where it could lead to. Keep in mind $|\Delta_{F_P}^{(4)}|\sim 0.1 F_P^2$ and $Y<1.1$ as suggestions following from \cite{SternRes, DescotesRes}. These assumptions hint the following  consequences, demonstrated in Fig.3

\vspace{-0.2cm}
\begin{itemize}
        \item[-] $r$$\,\sim\,$$15$ and
                 $\Delta_{F_{\eta}}^{(4)}$-$\Delta_{F_{\pi}}^{(4)}$$<$$0.2F_{\eta}^2$
                 implies $Y$$>$1 or $L_5^r(M_{\rho})$$>$$2.10^{-3}$.
                 \vspace{-0.25cm}
        \item[-] $r$$\,\sim\,$$25$ and
                 $\Delta_{F_{\eta}}^{(4)}$-$\Delta_{F_{\pi}}^{(4)}$$<$$0.2F_{\eta}^2$
                 implies $Y$$>$$0.5$ or $L_5^r(M_{\rho})$$>$$2.10^{-3}$.
                 \vspace{-0.25cm}
        \item[-] $L_5^r(M_\rho)$$<$$1.10^{-3}$ and $r$$\,\sim\,$$25$ implies  $Y$$>$1.2,
                 $\Delta_{F_{\eta}}^{(4)}$-$\Delta_{F_{\pi}}^{(4)}$$>$$0.2F_{\eta}^2$.
\end{itemize}
\vspace{-0.2cm}

\begin{figure}[t]
        \hspace{-0.5cm}
        \epsfig{figure=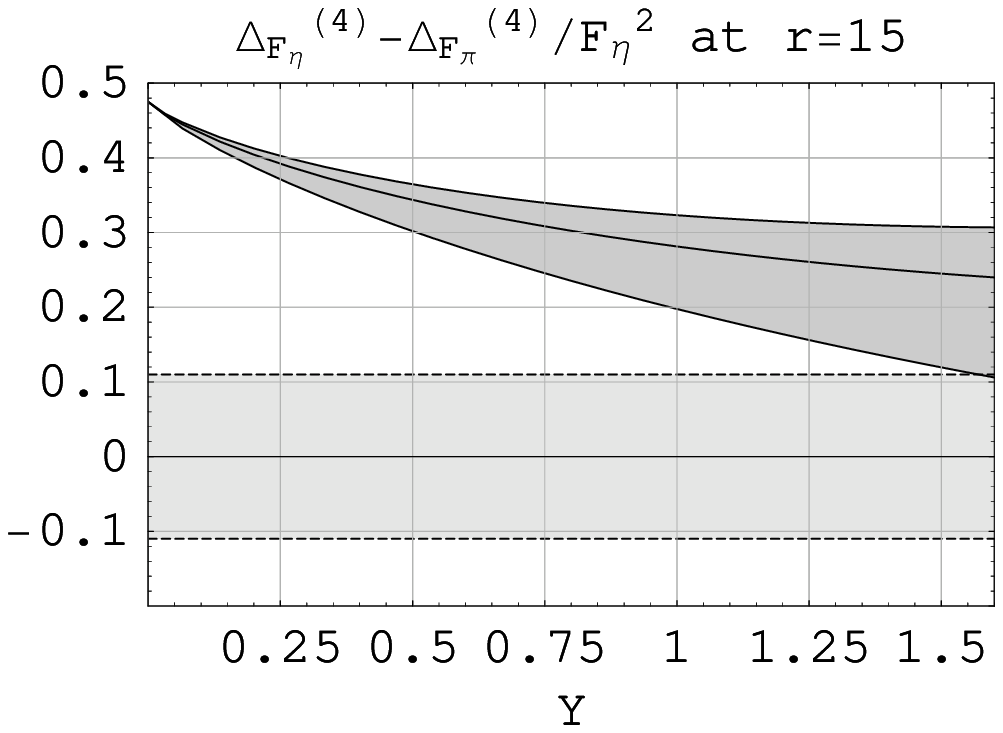,width=0.53\textwidth}
        \hspace{-0.5cm}
        \epsfig{figure=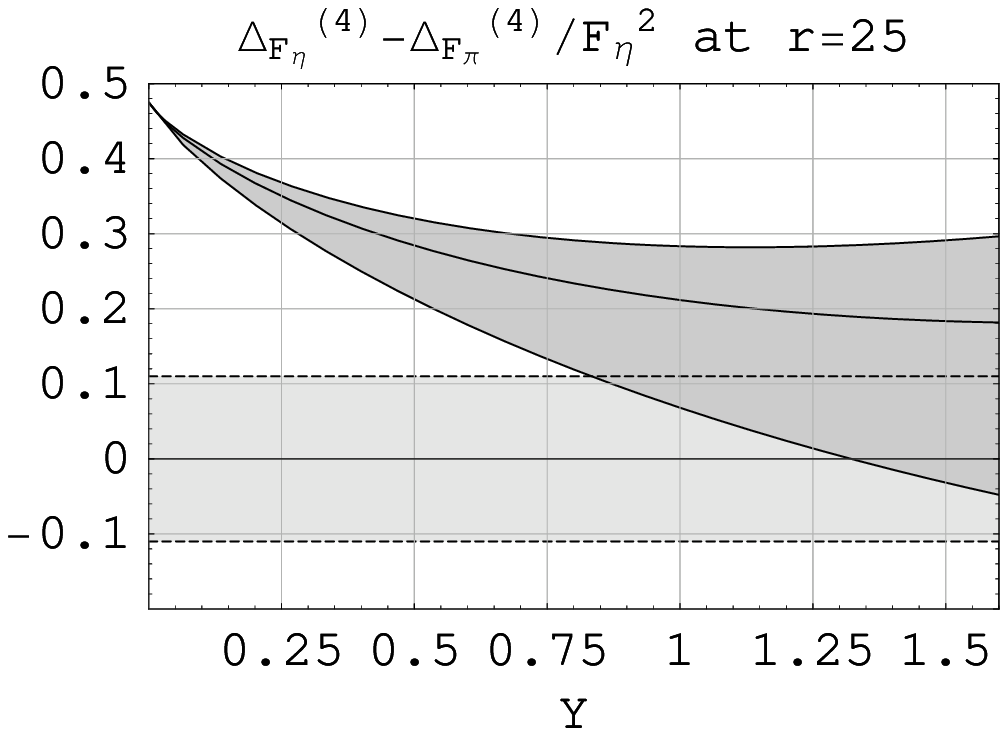,width=0.53\textwidth} \\ \\
        Figure 3: $\Delta_{F_{\eta}}^{(4)}$-$\Delta_{F_{\pi}}^{(4)}$ estimate for
                  $L_5^r(M_{\rho})\sim(0.5-2).10^{-3}$ (dark band). Upper bound corresponds to 
                  low values of $L_5^r$. Left: $r$=15, right: $r$=25. \\
                  Light: expected uncertainty $\pm0.11F_{\eta}^2$ from the 10\% uncertainty 
                  estimate.
\end{figure}

\noindent The remainder estimate using the Generalized $\chi$PT Lagrangian \cite{EtaPi} provides

\begin{myeqn}{0cm}
         3\Delta_{F_{\eta}}^{(4)}-4\Delta_{F_K}^{(4)}+\Delta_{F_{\pi}}^{(4)}\ =\
        2A_2^rF_{\pi}^2\hat{m}^2(r-1)^2 + 8A_3^rF_{\pi}^2\hat{m}^2(r^2+1)
        - 4B_2^r(\mu)F_{\pi}^2\hat{m}^2(r-1)^2
\next{1.5cm}
        -\ 8{C_1^P}^r(\mu)F_{\pi}^2\hat{m}^2(r-1)^2
        - \frac{M_{\pi}^2(X-1)}{16\pi^2}\ln[\frac{M_{\pi}^2}{\mu^2}]
\last{0cm}
        - \frac{4M_K^2-2M_{\pi}^2(r+1)X}{16\pi^2}\ln[\frac{M_K^2}{\mu^2}]
        + \frac{3M_{\eta}^2-M_{\pi}^2(2r+1)X}{16\pi^2}\ln[\frac{M_{\eta}^2}{\mu^2}]
        + \Delta_{G\chi PT}^{(5)}.
\end{myeqn}

We use two ways to estimate the unknown G$\chi$PT LEC's. The first is the usual simple variation of scale, the constants are set to zero at two different scales and the sensitivity is checked.  The second assumes a probabilistic distribution of possible values depending on scale variation

\begin{myeq}{0cm}
        B_2^r(M_{\rho})\, =\, 0 \pm \frac{Z_0^S+Z_0^P}{4\pi^2F_{\pi}^2}
                                \ln [\frac{1 \textrm{GeV}}{M_{\rho}}],\ \ \
        {C_1^P}^r(M_{\rho})\, =\, 0 \pm \frac{A_0-Z_0^S}{16\pi^2F_{\pi}^2}
                                    \ln [\frac{1 \textrm{GeV}}{M_{\rho}}].
\end{myeq}

\noindent These are then added in squares. Note that the insensitivity in the first case assures independence on where the central value is chosen in the latter one.

The results for $r$=25 can be seen in Fig.4, low values of $r$ do not change the overall picture. However, of the four G$\chi$PT LEC's present in our case only two depend on scale, which is hardly a good statistical ensemble. There is no indication of large higher order corrections nevertheless.

\begin{figure}
        \hspace{-0.5cm}
        \epsfig{figure=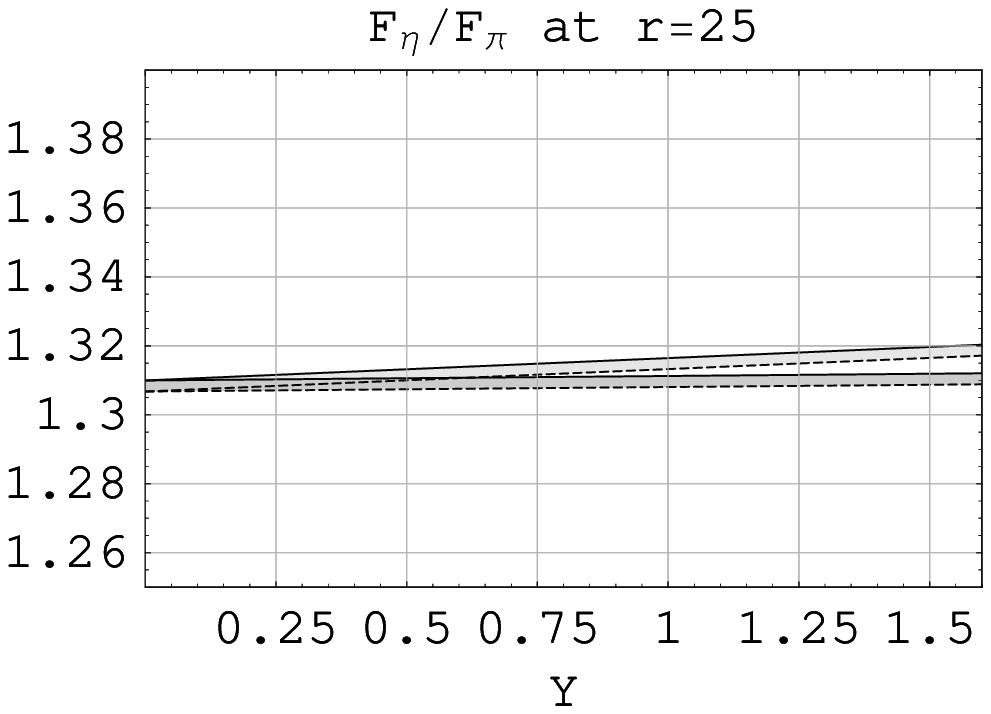,width=0.53\textwidth}
        \hspace{-0.5cm}
        \epsfig{figure=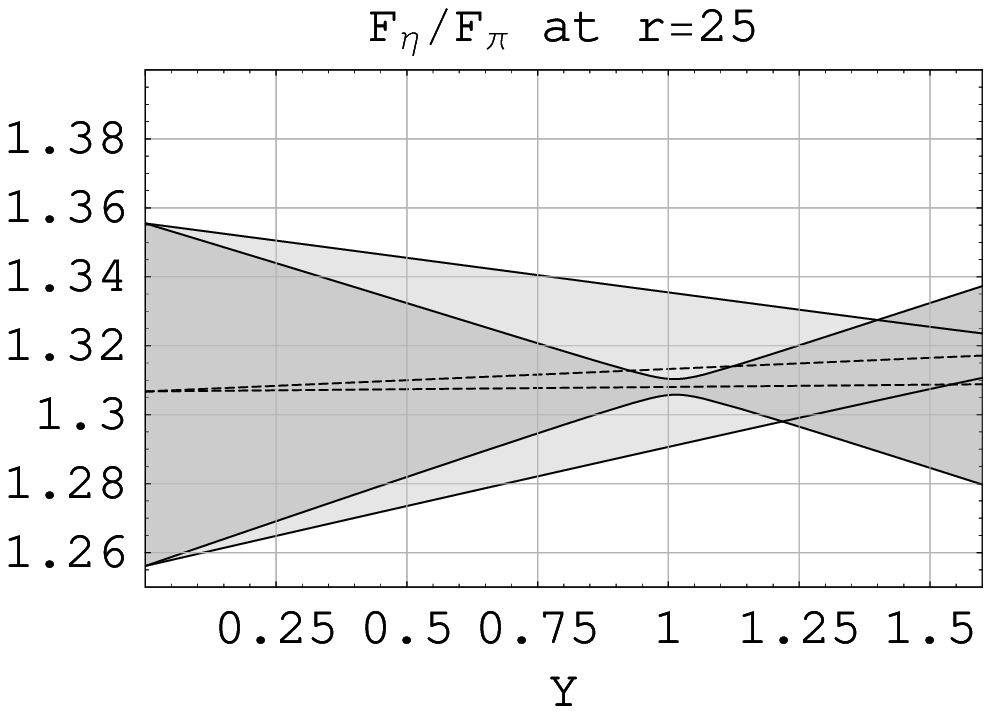,width=0.53\textwidth} \\ \\
        Figure 4: G$\chi$PT remainder estimate for $r$=25. Dark: $Z$=0.9, light: $Z$=0.5.
                  Left: simple variation of scale, solid: $\mu=1$GeV, dashed: $\mu=M_{\rho}$.
                  \newline Right: the LEC estimate described in the
                  text, solid: error bars $\mu=1\textrm{GeV}/M_{\rho}$, dashed: central values
                  $\mu=M_{\rho}$.
\end{figure}

\section{Summary}

We have studied the case of pseudoscalar decay constants in the 'Resummed' $\chi$PT framework and tried to obtain an estimate for the $\eta$ decay constant and related parameters. We used several ways to get a feeling about the effect of higher order remainders.
\\ \\
Acknowledgment: This work was supported in part by the Center for Particle Physics
(project no. LC 527) and by the EU Contract No. MRTN-CT-2006-035482, \lq\lq FLAVIAnet''.

\end{document}